\documentclass[11pt]{article}
\usepackage{moriond,epsfig}
\usepackage{amsmath}
\usepackage{graphicx}
\usepackage{amsfonts}
\usepackage{amssymb}
\usepackage{caption}
\usepackage{natbib}

\bibpunct{[}{]}{,}{n}{,}{,}

\def\be{\begin{equation}}
\def\ee{\end{equation}}
\def\bea{\begin{eqnarray}}
\def\eea{\end{eqnarray}}

\begin{document}
\vspace*{4cm}
\title{THEORETICAL PERSPECTIVE ON THE NA62 PHYSICS PROGRAM}

\author{CHRISTOPHER SMITH }

\address{Université Lyon 1 \& CNRS/IN2P3, UMR5822 IPNL\\
4 rue Enrico Fermi, 69622 Villeurbanne Cedex, France}

\maketitle\abstracts{
Soon the NA62 experiment will start looking for the rare $K^+\rightarrow \pi^+\nu\bar\nu$ decay. In this talk, its theoretical interests, together with those of the neutral rare decays $K_L\rightarrow \pi^0\nu\bar\nu$, $K_L\rightarrow \pi^0 e^+e^-$, and $K_L\rightarrow \pi^0\mu^+\mu^-$, are briefly reviewed. Then, other possible targets for NA62 are discussed, among which the dominant semileptonic decays, the radiative decays, as well as the lepton-flavor violating decays.}

\section{New physics searches}

The rare $K$ decays are ideally suited to search for New Physics (NP). Besides the loop suppression of the underlying FCNC processes, they are significantly suppressed by the CKM matrix elements. Indeed, compared to the $b\rightarrow s,d$ transitions, they scale dominantly as
\begin{equation}
s\rightarrow d\sim|V_{td}V_{ts}^{\dagger}|\sim\lambda^{5}\;\;\ll
\;\;b\rightarrow d\sim|V_{td}V_{tb}^{\dagger}|\sim\lambda^{3},\;\;b\rightarrow
s\sim|V_{ts}V_{tb}^{\dagger}|\sim\lambda^{2}\;,\label{SMCKM}%
\end{equation}
with $\lambda\approx0.22$. So, it is in $K$ physics that NP is most easily competitive with the Standard Model (SM). For example, some NP effects can be encoded into the following effective operators
\begin{equation}
\mathcal{H}_{eff}=\frac{c_{sd}}{\Lambda^{2}}(\bar{s}\Gamma d)(\bar{\nu}%
\Gamma\nu)+\frac{c_{bd}}{\Lambda^{2}}(\bar{b}\Gamma d)(\bar{\nu}\Gamma
\nu)+\frac{c_{bs}}{\Lambda^{2}}(\bar{b}\Gamma s)(\bar{\nu}\Gamma\nu)\;,
\end{equation}
with $\Gamma$ some Dirac structure. Clearly, if this NP is generic, $c_{ij}=\mathcal{O}(1)$, the constraints from $K\rightarrow \pi\nu\bar\nu$ are potentially the toughest. A measurement close to the SM predictions would require $\Lambda>\mathcal{O}(100 \text{ TeV})$. Even accounting for possible model-dependent loop and gauge couplings, this measurement would be the most difficult to reconcile with the existence of generic NP at a relatively low scale. This is one instance of the so-called NP flavor puzzle.

Alternatively, the NP model could preserve the CKM scalings (\ref{SMCKM}), i.e. $c_{ij}=\mathcal{O}(|V_{tj}V_{ti}^{\dagger}|)$. Such models are referred to as satisfying Minimal Flavor Violation~\cite{MFV} (see also~\cite{MFValt}). When this is the case, it could show up at a scale $\Lambda\lesssim1$ TeV without violating the experimental bounds. In addition, when MFV is enforced within the MSSM, the effects on the rare $K$ decays are expected to be small, beyond the experimental sensitivity. This has been analyzed at moderate~\cite{IMPST} and large $\tan\beta$~\cite{LargeB,LargeBnu} (with $\tan\beta\equiv v_{u}/v_{d}$ and $v_{u,d}$ the two MSSM Higgs vacuum expectation values), without R-parity~\cite{MFVRPV}, or with MFV imposed at the GUT scale~\cite{MFVRGE}. Turning this around, the rare $K$ decays are one of the best places to look for deviations with respect to MFV. If the flavor-breaking transitions induced by NP are not precisely aligned with those of the SM (tuned by the CKM), significant effects could show up.

\subsection{Where do we stand and where do we go?}

\begin{figure}[t]
\centering \includegraphics[width=0.85\textwidth]{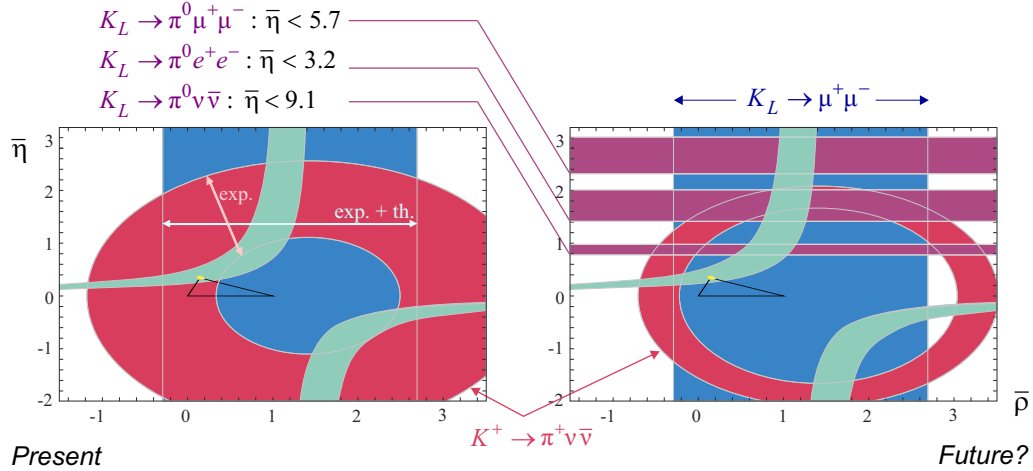}
\captionsetup{justification=justified,font=small}
\caption{Rare $K$ decays in the $\bar{\rho}-\bar{\eta}$ plane: the situation now and in a hypothetical future. For the current situation, the experimental inputs are $\mathcal{B}(K^+\rightarrow \pi^+\nu\bar\nu)=17.3_{-10.5}^{+11.5}\cdot10^{-11}$ ~\cite{EKP}, $\mathcal{B}(K_L\rightarrow \pi^0\nu\bar\nu)<2.6\cdot10^{-8}$ ~\cite{EK0}, $\mathcal{B}(K_L\rightarrow \pi^0e^+e^-)<28\cdot10^{-11}$ ~\cite{KTEVe}, and $\mathcal{B}(K_L\rightarrow \pi^0\mu^+\mu^-)<38\cdot10^{-11}$ ~\cite{KTEVm}. For the latter two modes, the bounds are derived without any assumption on the sign of the interference between the direct and indirect CP-violating contributions (see Ref.~\cite{MesciaST} for more details). For the future situation, the 90\% confidence regions are obtained by taking the measured central values as $\mathcal{B}(K^+\rightarrow \pi^+\nu\bar\nu)=17.3\cdot 10^{-11}$, $\mathcal{B}(K_L\rightarrow \pi^0\nu\bar\nu)=25\cdot10^{-11}$, $\mathcal{B}(K_L\rightarrow \pi^0e^+e^-)=17\cdot10^{-11}$, and $\mathcal{B}(K_L\rightarrow \pi^0\mu^+\mu^-)=12\cdot10^{-11}$, each of them with a $1\sigma$ experimental error of $15\%$, and including all the current theoretical errors but for the interference sign between the direct and indirect CP-violating contributions to the $e^+e^-$ and $\mu^+\mu^-$ modes, assumed positive.}
\label{fig:CKM2}%
\end{figure}

To test the CKM picture of the SM, it is customary to represent the constraints coming from flavor physics in the $\bar{\rho}-\bar{\eta}$ plane~\cite{CKMfits}. As shown in Fig.~\ref{fig:CKM2}, rare $K$ decays are currently not competitive compared to $B$ physics observables. Furthermore, though mostly experimentally driven, this situation should persist for the foreseeable future.

However, this way of presenting the impact of $K$ physics for the search for NP is not doing justice to its true potential. The main reason why rare $K$ decays are not competitive is their significant CKM suppression compared to typical $B$ physics observables. But, as said previously, being so suppressed leaves more chances for NP to be competitive. Sizeable deviations with respect to the SM could easily show up. This is illustrated by the second plot in Fig.~\ref{fig:CKM2}, where an imaginary future situation is drawn. Despite the rather large error bands standing for both the whole theoretical errors and the hypothesized future 15\% experimental errors, the deviations could be sufficient to clearly signal non-standard physics. Importantly, this NP needs not affect all the rare $K$ decays equally. The global pattern of deviations would be an important piece of information to reconstruct the NP at play, as we will now discuss.

\subsection{How to disentangle New Physics effects}

The first target of NA62 is the $K^{+}\rightarrow\pi^{+}\nu\bar{\nu}$ decay, induced by the $Z$ penguin and its associated $W$ box. Technically, this FCNC has a peculiar sensitivity to the high-scale because it would be represented by a dimension-six operator if $SU(2)_{L}$ was exact. Instead, after the spontaneous breakdown of $SU(2)_{L}$, the operator $\bar{s}_L\gamma_{\mu}d_LH^{\dagger}D^{\mu}H$ becomes $v^{2}\bar{s}_L\gamma_{\mu}d_L Z^{\mu}$, which is effectively of dimension four and enhanced by two powers of the electroweak vacuum expectation value $v\approx250$ GeV. Most NP models affect the $Z$ penguin (or lead to an indistinguishable $V\otimes(V-A)$ operator). This has  been extensively studied, so let us just cite, besides the analyses within MFV quoted previously, the MSSM at moderate $\tan\beta$ (chargino loops)~\cite{IMPST,Chargino}, MSSM at large $\tan\beta$ (charged Higgs loops)~\cite{LargeBnu}, R-parity violation (non MFV)~\cite{RPV}, enhanced electroweak penguins~\cite{EEP}, Little Higgs~\cite{LH}, extra dimensions~\cite{ED}, fourth generation~\cite{4G},...

\begin{figure}[t]
\centering       \includegraphics[width=0.45\textwidth]{Fig2.eps}
\ \ \includegraphics[width=0.49\textwidth]{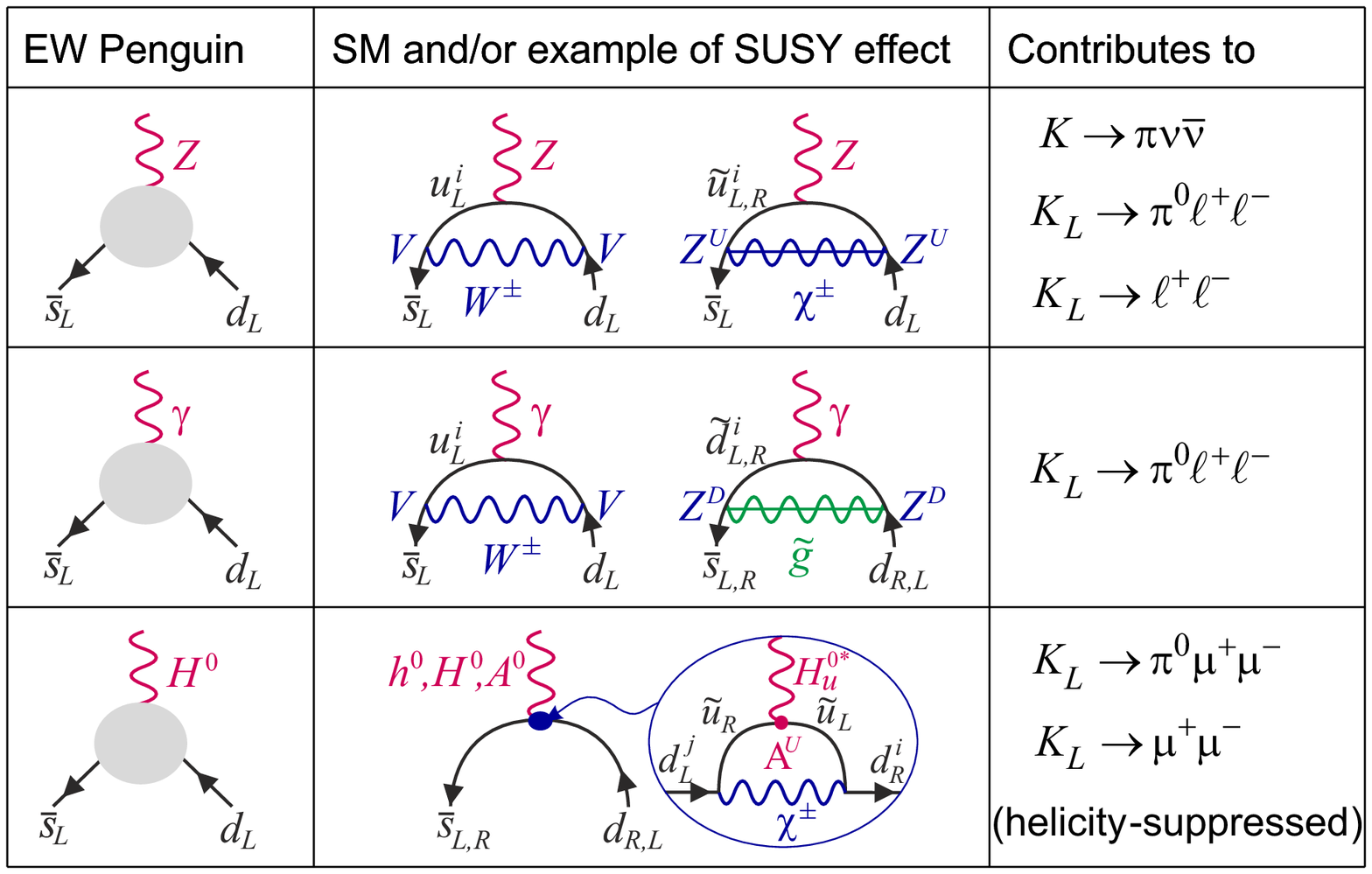}  
\captionsetup{justification=justified,font=small}
\caption{The discriminating power of (left) the pair of decays $K^{+}\rightarrow\pi^{+}%
\nu\bar{\nu}$--$K_{L}\rightarrow\pi^{0}\nu\bar{\nu}$, and of (right) the full set of rare $K$ decays, with examples of effects in the MSSM. }%
\label{fig:DP}%
\end{figure}

Since the $K^{+}\rightarrow\pi^{+}\nu\bar{\nu}$ rate is just one number, we need more information to disentangle all these models. A first clue would be provided by the CP-violating $K_{L}\rightarrow\pi^{0}\nu\bar{\nu}$ mode. Model-independently~\cite{GN}, the current measurement of $K^{+}\rightarrow \pi^{+}\nu\bar{\nu}$ allows for up to a factor $30$ enhancement of $K_{L}\rightarrow\pi^{0}\nu\bar{\nu}$ with respect to its SM value. The discriminating power of the pair of $K\rightarrow\pi\nu\bar{\nu}$ decays is shown in Fig.~\ref{fig:DP}, where the grid in the allowed region is a function of the theoretical errors on their SM predictions.

In general, the NP decouple smoothly as its mass scale increases or its flavor-breaking couplings decrease, so there is naturally a cluster of models around the SM values in Fig.~\ref{fig:DP}. Thus, in case the deviations are not large, information beyond $K\rightarrow\pi\nu\bar{\nu}$ will be needed to identify the NP at play. To this end, one should turn to the $K_{L}\rightarrow\pi^{0}\ell^{+}\ell^{-}$ ($\ell=e,\mu$) decays. Though less clean than $K\rightarrow\pi\nu\nu$ theoretically (see next section), their sensitivity to NP is significantly different~\cite{MesciaST} because $\ell^{+}\ell^{-}$ not only couple to the $Z$, but also to the $\gamma$ and, for $\ell=\mu$, to the Higgs(es). In Fig.~\ref{fig:DP} are examples of mechanism within the MSSM affecting these various electroweak structures. Going back to Fig.~\ref{fig:CKM2}, the pattern depicted would be interpreted as indicating NP in the $Z$ penguin, since $K\rightarrow\pi\nu\bar{\nu}$ disagree with the SM, in the $\gamma$ penguin since $K_{L}\rightarrow\pi^{0}e^{+}e^{-}$ disagrees even more, and in the Higgs penguins because $K_{L}\rightarrow\pi^{0}\mu^{+}\mu^{-}$ is yet more affected than $K_{L}\rightarrow\pi^{0}e^{+}e^{-}$ (see Fig.~4 in Ref.~\cite{MesciaST}).

\section{Hadronic uncertainties and Chiral Perturbation Theory}

\begin{figure}[t]
\centering       \includegraphics[width=0.52\textwidth]{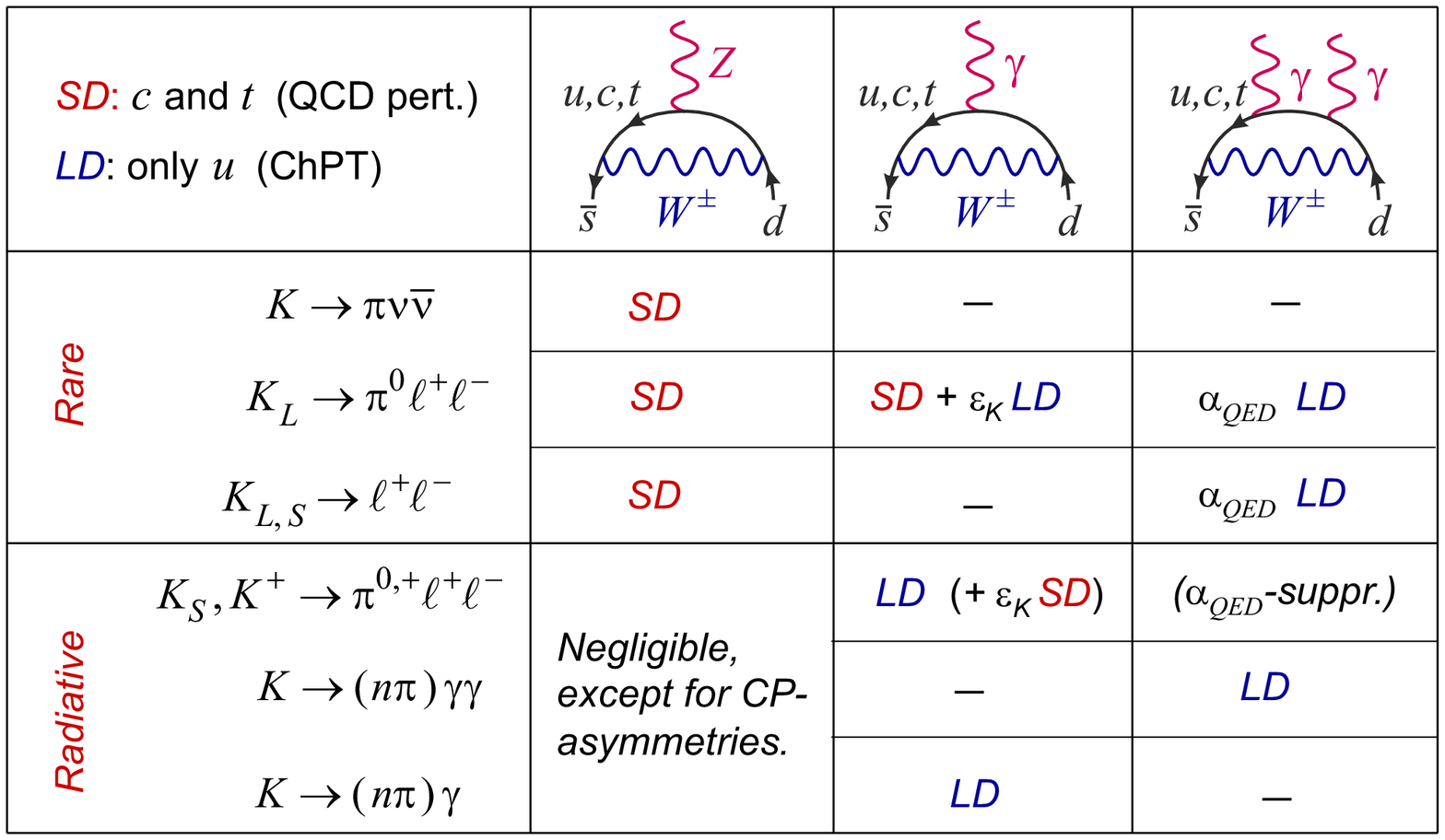}
\includegraphics[width=0.44\textwidth]{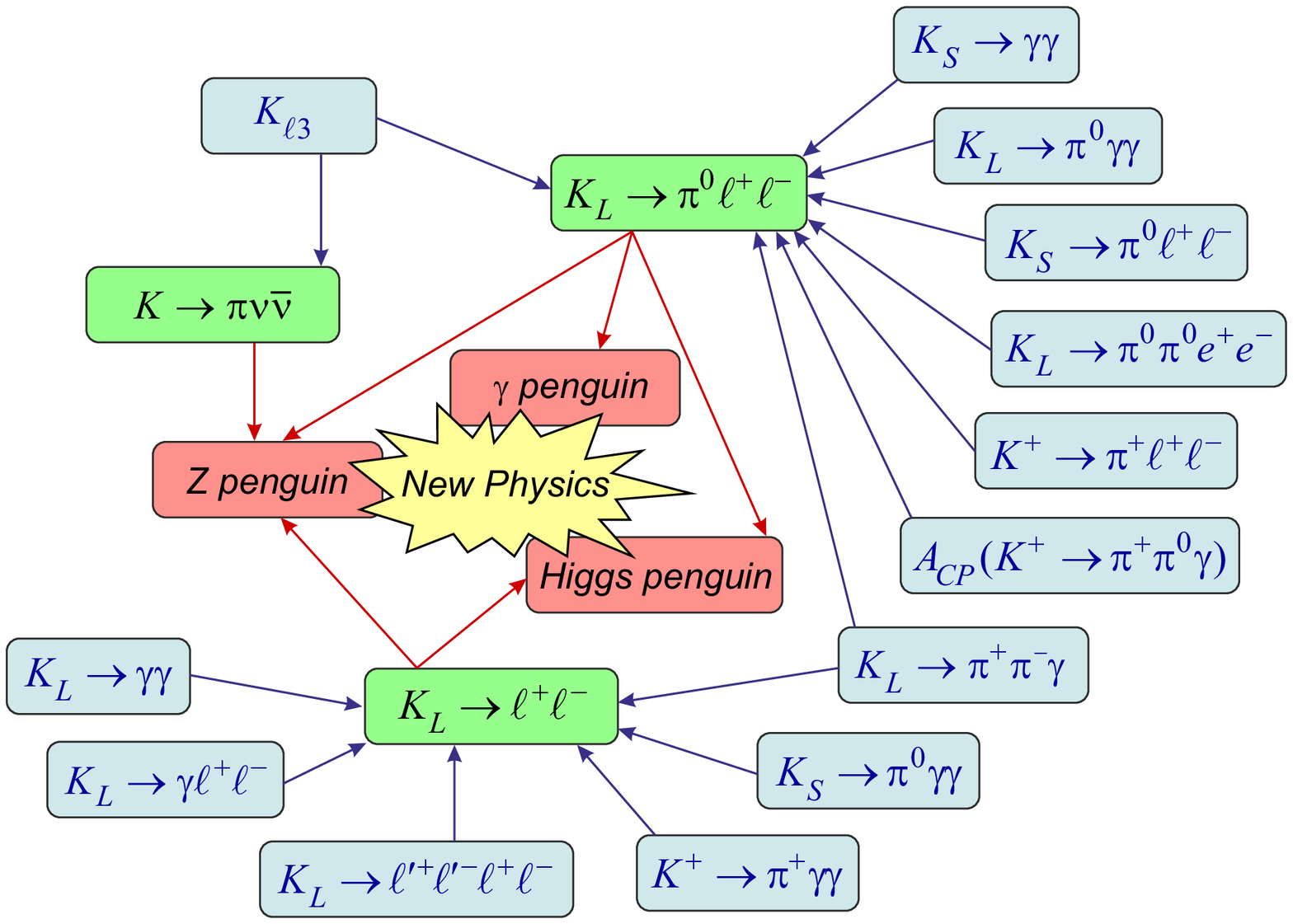}  
\captionsetup{justification=justified,font=small}
\caption{Anatomy of the rare $K$ decays in the SM, and strategies to fix their LD uncertainties.}\label{fig:21}%
\end{figure}

The rare $K$ decays proceed dominantly through Short-Distance (SD) processes, with only residual Long-Distance (LD) effects, see Fig.~\ref{fig:21}. Still, to make the most from future measurements, control over these effects is compulsory. To this end, the strategy is to use Chiral Perturbation Theory (ChPT) to relate the hadronic uncertainties occurring for rare $K$ decays to other observables, which thus constitute important secondary targets for NA62. We can distinguish two kinds of LD effects. The first are the matrix elements of the local semileptonic effective operators induced by the short-distance physics (i.e., penguins with top and charm quarks), and the second are the genuine non-local LD processes (i.e., penguins with up quarks).

For the matrix elements, all the necessary information can be extracted from the charged-current semileptonic $K\rightarrow\pi\ell\nu$ decays (called $K_{\ell3}$). In the isospin limit, their hadronic matrix elements $\langle\pi|\bar{s}\gamma_{\mu}u|K\rangle$ are equal to those of the neutral current, $\langle\pi|\bar{s}\gamma_{\mu}d|K\rangle$, up to some Clebsch-Gordan coefficients. Within ChPT, the isospin breaking effects can be brought under control, ensuring a few per-mil accuracy for the matrix elements~\cite{Matrix}. This represents a tiny fraction of the overall error on the SM predictions for the $K\rightarrow\pi\nu\bar{\nu}$ rates~\cite{Kpnn}, and could even be further improved with better measurements of the $K_{\ell3}$ rates and slopes:%
\begin{align}
B(K^{+}\overset{}{\rightarrow}\pi^{+}\nu\bar{\nu}(\gamma))^{\text{SM}} &
=8.2(8)\cdot10^{-11}\;\;\;\;(53_{CKM},31_{SD},14_{LD},2_{ME})\%\;,\\
B(K_{L}\overset{}{\rightarrow}\pi^{0}\nu\bar{\nu})^{\text{SM}} &
=2.6(4)\cdot10^{-11}\;\;\;\;(84_{CKM},14_{SD},2_{ME})\%\;,
\end{align}
where in the breakdown of the errors, $CKM$ indicates parametric errors, and $ME$ those from the matrix elements~\cite{Stamou}. In this respect, we should mention that the $K_{\ell3}$ decays are also essential to extract~\cite{Flavia} $V_{us}$, which enters in the parametric uncertainty.

The second kind of LD effects occurs predominantly for the charged lepton modes, where they are due to the $\gamma$ and $\gamma\gamma$ penguins, see Fig.~\ref{fig:21}. These FCNC, contrary to the $Z$ penguin, do not suppress light-quark contributions. It is only thanks to the CP symmetry that the bulk of the LD contributions is projected out for $K_{L}\rightarrow\pi^{0}\ell^{+}\ell^{-}$. To estimate the remaining LD effects, one can use those modes in which the $\gamma$ or $\gamma\gamma$ penguin is CP-conserving, and represent the dominant contributions. We will not detail here the strategies depicted in Fig.~\ref{fig:21}, and instead refer to some recent theoretical works along
these lines~\cite{Kpll}. The main message is that all these radiative modes should be tackled by NA62. Not only are they interesting by themselves to study the interplay between the strong, weak, and electromagnetic interactions (including the QED anomaly), but they are also an important ingredient in the search for NP.

\section{Precision physics}

The NA62 experiment aims at producing about $10^{13}$ kaons per year. At that level, one is clearly entering a high-precision era in kaon physics. Thus, besides the rare $K$ decays, there are many observables worth including in the physics program. Of course, to really appreciate if an observable is worth the experimental effort, one should first reassess whether theory can match the unprecedented precision, or at least whether the theoretical control is sufficient to access to some interesting physics. Though this work is currently in progress~\cite{Handbook}, let us mention a few possible targets.

\subsection{CP-violation}

The dominant $K\rightarrow\pi\pi$ decays permit to exact the indirect CP-violating parameter $\varepsilon_{K}$, due to the $\Delta S=2$ mixing, and the direct CP-violating parameter $\varepsilon^{\prime}$, due to $\Delta S=1$ penguins. Both are already well known experimentally, and the ball is in the theorists' camp. While lattice~\cite{MesciaProc} has a good prospect at reaching a percent level precision for the matrix elements of the effective $\Delta S=2$ light-quark operator, problems remain for $\varepsilon^{\prime}$. The question is thus whether NA62 could help in this respect.

There are several other windows into direct CP-violation, the first of which being the rare $K$ decays discussed previously. But besides these, one could also turn to direct CP-asymmetries,%
\begin{equation}
A_{CP}=\frac{\Gamma(K^{+}\rightarrow X)-\Gamma(K^{-}\rightarrow\bar{X}%
)}{\Gamma(K^{+}\rightarrow X)+\Gamma(K^{-}\rightarrow\bar{X})}\;,\;\;X=\pi
\pi\pi,\;\pi\pi\ell\nu,\;\pi\pi\gamma,\;\pi\gamma\gamma,\;\pi\ell\bar{\ell
},...\;,
\end{equation}
as studied for example in Refs.~\cite{MesciaST,Kpll,DAmbrosioI96,Frere}. Of course, these asymmetries are in general very small, but they have specific sensitivities to scalar, electroweak, and/or QCD penguins. They would provide complementary information with respect to the rare $K$ decays. The main issues for NA62 are first to run with both $K^{-}$ and $K^{+}$ beams, and then to be able to reach asymmetries in the $10^{-4}-10^{-5}$ range. If $K^{-}$ beams are not practical, phase-space asymmetries for a given $K^{+}$ decay mode could also help. Note that the neutral $K$ decays are in general less clean because indirect CP-violation is dominant; i.e. a CP-asymmetry essentially gives back the already well-measured $\varepsilon_{K}$, relegating direct CP-violating effects to small corrections.

\subsection{Other subjects under investigation}

There are many other possible targets for NA62, which we can organize in two categories. From them, either we would learn more about QCD in the non-perturbative regime, or we would directly constrain NP. In the former case, improving our theoretical control on QCD effects leads indirectly to better NP constraints, by allowing us to fine-tune our theoretical tools like ChPT or Lattice QCD.

A first set of targets are the leptonic and semileptonic decays $K_{\ell2}$ and $K_{\ell3}$, which provide very delicate tests of the SM. We have discussed earlier how $K_{\ell3}$ data permits to fix rare $K$ decay matrix elements. In addition, as presented at this conference, these modes also permit to test leptonic universality~\cite{Univers} and the CKM unitarity~\cite{Unitar}. Furthermore, their measurements allow for high precision studies of the charge-current form-factors, including isospin breaking effects, and could provide one of the best sources of information on the light-quark mass ratios~\cite{Leutwyler}.

Other modes of interests are the lepton-flavor violating decay channels $K\rightarrow(\pi)\mu e$. Those are forbidden in the SM, so any event would unambiguously signal the presence of NP. As such, they should certainly be included in the NA62 physics program, even though in most (but not all~\cite{LQ}) models, they are correlated with purely leptonic observables like $\mu\rightarrow e\gamma$, and thus already severely constrained~\cite{LFV}.

Finally, we should mention the $K_{\ell4}$ decays as well as the hadronic decays giving us access to the $\pi\pi$ scattering phases, which have been the focus of intense theoretical work~\cite{pipi}. Also, pion and hyperon decays are accessible to NA62, which could complement or improve the experimental situation in some channels.

\section{Conclusion}

The upcoming NA62 experiment at CERN as well as the K0TO experiment in Japan are perfectly positionned to unravel the still elusive NP. First, their main targets, the rare $K\rightarrow\pi\nu\bar{\nu}$ decays, are under excellent control theoretically. The non-parametric errors on the SM predictions for their rates are well below $10\%$. Second, the impact of various NP scenarios have been extensively studied, and usually found significant. Taken in combination with the rare $K_{L}\rightarrow\pi^{0}\ell^{+}\ell^{-}$ decays, they will permit a detailed analysis of the $s\rightarrow d$ transition. This is an essential ingredient in the LHC era if one aims at reconstructing the flavor sector of the NP at play~\cite{CERN}.

\section*{Acknowledgments}

The initial version of this work, published in the Moriond Proceedings, was supported by the European Commission RTN network, Contract No. MRTN-CT-2006-035482 \textit{FLAVIAnet}, and by project C6 of the DFG Research Unit SFT-TR9 \textit{Computergest\"{u}tzte Theoretische Teilchenphysik}.

\end{document}